\documentclass[runningheads]{llncs}

\usepackage{graphicx}
\usepackage{amssymb}
\usepackage{xcolor}
\usepackage{hyperref}
\usepackage{csquotes}  
\usepackage{booktabs}
\usepackage{balance} 
\usepackage{amssymb}
\usepackage{subcaption}
\usepackage{tablefootnote}
\usepackage[sort,square,numbers]{natbib}
\usepackage[indentfirst=false,font={itshape}, begintext=``,endtext='']{quoting}

\makeatletter
\newcommand{\ygg@basicalert}[2]{\fbox{\bfseries\sffamily\scriptsize#1}{\sf\small$\blacktriangleright$\color{red}{\textit{#2}}$\blacktriangleleft$}}
\newcommand{\GABRIEL}[1]{\ygg@basicalert{GABRIEL}{#1}}
\newcommand{\CRIS}[1]{\ygg@basicalert{CRISTIANO}{#1}}
\newcommand{\YANN}[1]{\ygg@basicalert{YANN}{#1}}
\newcommand{\FABIO}[1]{\ygg@basicalert{FABIO}{#1}}

\begin{document}

\title{Visualising Game Engine Subsystem Coupling Patterns}

 \author{
 Gabriel C. Ullmann\inst{1}\orcidID{0000-0002-3274-0789} \and
 Yann-Ga\"el Gu\'{e}h\'{e}neuc\inst{1}\orcidID{0000-0002-4361-2563} \and
 Fabio Petrillo\inst{2}\orcidID{0000-0002-8355-1494} \and
 Nicolas Anquetil\inst{3}\orcidID{0000-0003-1486-8399} \and
 Cristiano Politowski\inst{2}\orcidID{0000-0002-0206-1056} \\
 }

 \authorrunning{G. Ullmann et al.}

 \institute{Concordia University, Montreal QC, Canada \\
 \email{g\_cavalh@live.concordia.ca, yann-gael.gueheneuc@concordia.ca}\\
 \and
 École de Technologie Supérieure, Montreal QC, Canada \\
 \email{fabio.petrillo@etsmtl.ca, cristiano.politowski@etsmtl.ca}
 \and
 Univ. Lille, CNRS, Inria, Centrale Lille, UMR 9189 - CRIStAL, Lille, France\\
 \email{nicolas.anquetil@inria.fr}}
\maketitle

\begin{abstract}
    Game engines support video game development by providing functionalities such as graphics rendering or input/output device management. However, their architectures are often overlooked, which hinders their integration and extension. In this paper, we use an approach for architecture recovery to create architectural models for 10 open-source game engines. We use these models to answer the following questions: Which subsystems more often couple with one another? Do game engines share subsystem coupling patterns? We observe that the Low-Level Renderer, Platform Independence Layer and Resource Manager are frequently coupled to the game engine Core. By identifying the most frequent coupling patterns, we describe an emergent game engine architecture and discuss how it can be used by practitioners to improve system understanding and maintainability.
\end{abstract}

\section{Introduction}
\label{sec:intro}
Game engines are tools made to support video game development. From the perspective of Software Engineering, game engines are systems composed of subsystems, each providing functionalities essential for any video game, such as 2D/3D graphics rendering or input/output device management. However, the versatility of game engines also makes them architecturally complex and often difficult to understand. The lack of architecture understanding hinders software integration and extension, which is important in the context of plugin-extendable game engines such as Unreal, Unity and Godot. Therefore, studying game engine architecture is necessary: ``[a] prerequisite for integration and extension is the comprehension of the software. To understand the architecture, we should identify the architectural patterns involved and how they are coupled.'' \cite{agrahari_whats_2021}.

In this paper, we apply the approach for game engine architecture recovery described in our previous paper \cite{ullmann_gas_2023} to 10 popular open-source game engines. By following this approach, we obtain \textit{include} graphs tagged by subsystem, which we call architectural models for the sake of simplicity, for each game engine. 

By studying these models' nodes and relationships, we answer the following research questions:

\begin{itemize}
\item \textbf{RQ1}: Which subsystems more often couple with one another? 
\item \textbf{RQ2}: Do game engines share subsystem coupling patterns?
\end{itemize}

The remainder of the paper is organized as follows: Section \ref{sec:related_work} presents related work on game engine architecture and architectural recovery. Section \ref{sec:approach} describes our game engine architecture recovery approach. Section \ref{sec:results} shows the architectural models resulting from applying our approach and Section \ref{sec:discussion} discusses lessons learned from frequent coupling patterns. Section \ref{sec:threats} presents threats to validity and Section \ref{sec:conclusion} concludes with future work.


\section{Related Work}
\label{sec:related_work}

\begin{figure}[ht]
    \centering
    \includegraphics[width=0.9\textwidth]{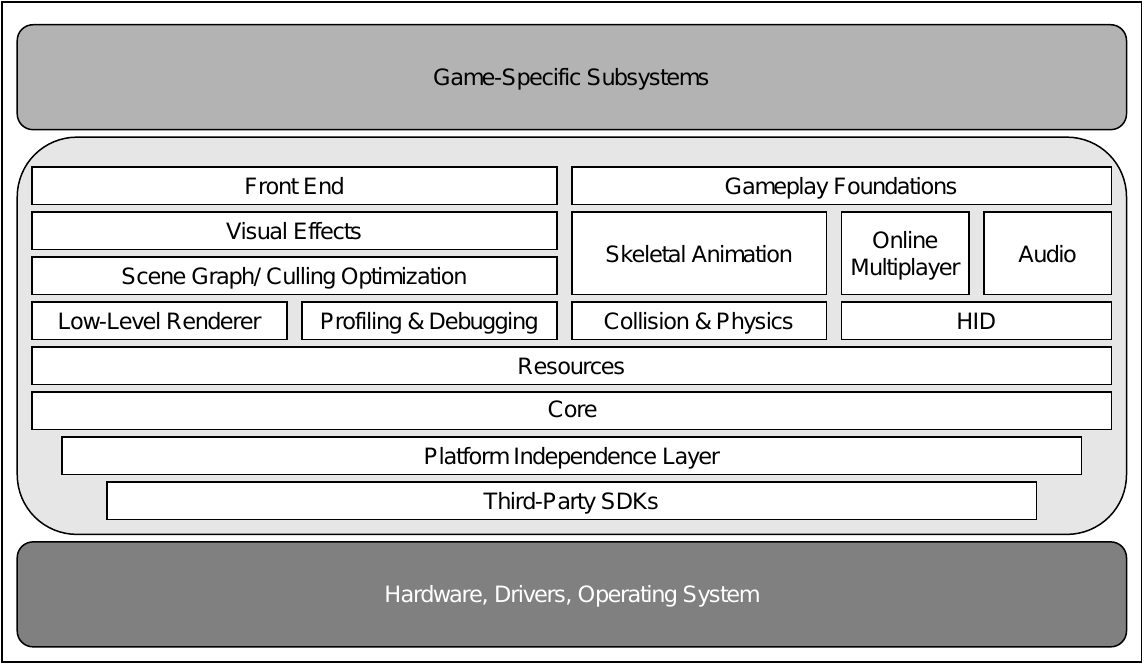}
    \caption{Runtime Engine Architecture, adapted from \citeauthor{gregory_game_2018} \cite[p.~33]{gregory_game_2018}.}
    \label{fig:gregory_arch}
\end{figure}

Few studies focused solely on game engine architecture and those that did focused on describing a specific game engine in detail. For example, \citeauthor{jaeyong_park_development_1997} \cite{jaeyong_park_development_1997} explain the scripting and networking subsystems of a game engine for MUDs \footnote{Multi-User Dungeon, a text-based precursor of MMORPG games.}. \citeauthor{bishop_designing_1998} \cite{bishop_designing_1998} describe the NetImmerse engine. Both authors represent subsystems and their relationships in diagrams. However, they do not discuss why these relationships exist, how often they appear or whether they are representative of all game engine architectures. 

\citeauthor[p.~33]{gregory_game_2018} \cite{gregory_game_2018} proposed a ``Runtime Game Engine Architecture'' (\autoref{fig:gregory_arch}), which describes common subsystems, their responsibilities, and some of their relationships. While also not focused on subsystem relationships, this is, to the best of our knowledge, the most comprehensive game engine architecture. Therefore we use it as a reference architecture in the following, as we explain further in \autoref{sec:subsystem_selection}.

Guided by a reference architecture, we apply an architecture recovery approach. Architecture recovery is concerned with the extraction of architectural descriptions from a system implementation \cite{bowman_linux_1999}. Researchers have applied it to systems such as the Apache Web server \cite{hassan_reference_2000}, the Android OS and Apache Hadoop \cite{link_value_2019} as a way to improve understanding and maintainability.

Game engine developers can also reap the benefits of architecture recovery for their systems. For example, the information obtained through architecture recovery can be ``used in the process of identifying suitable improvements and enhancements to a specific engine and have supported implementing these in an appropriate manner'' \cite{munro_architectural_2009}. This potential for assisting software architectural improvement is the main reason why we chose to apply architectural recovery to game engines in this work.

\section{Approach}
\label{sec:approach}

\begin{figure}[ht!]
    \centering
    \includegraphics[width=\textwidth]{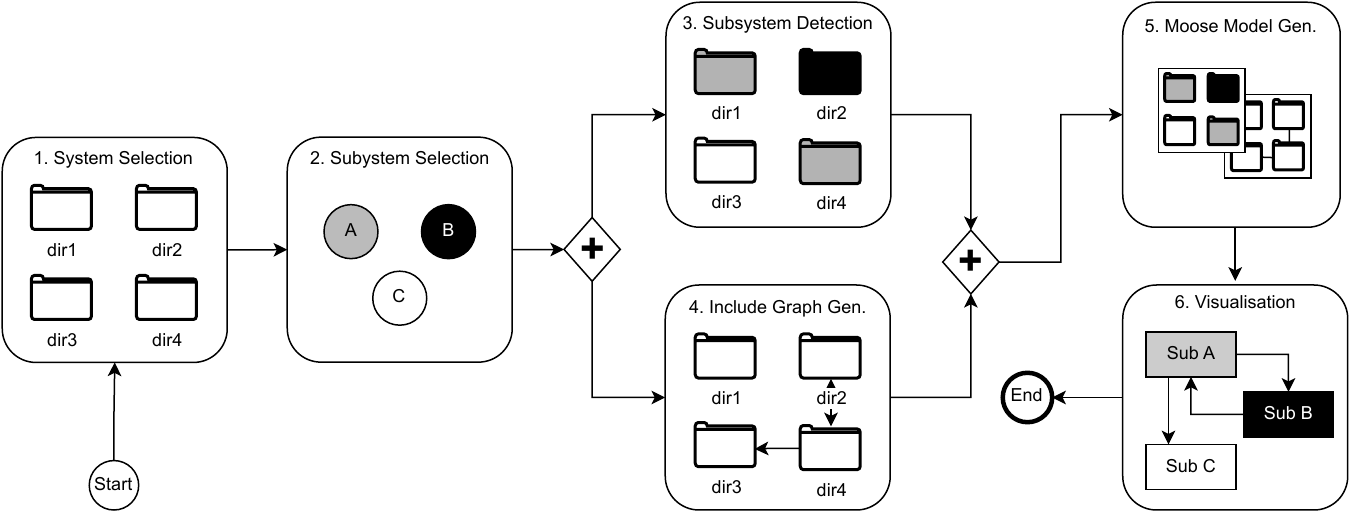}
    \caption{Steps of our architecture recovery approach.}
    \label{fig:approach}
\end{figure}

We now explain the six-step approach we proposed in a previous work \cite{ullmann_gas_2023}, summarized in \autoref{fig:approach} and used in this paper in a slightly modified version. Steps 1 to 3 are performed manually. Steps 4 to 6 are largely automated with Smalltalk and Python code, which is available on GitHub.\footnote{\url{https://github.com/gamedev-studies/game-engine-analyser}}

\subsection{System Selection}
\label{sec:sys-selection}

We searched for the term ``game engine'' on GitHub and then selected all repositories showing C++ as the predominant programming language, given its relevance to game engine development \cite{politowski_are_2021}. This initial selection consisted of 20 game engine repositories. We then removed from the selection all non-general-purpose game engines. For example, we did not select the engine \textit{minetest} \footnote{\url{https://github.com/minetest/minetest}} because it is limited to creating games in the style of Minecraft. Finally, we sorted the remaining repositories by the sum of their GitHub forks and stars (as of May 2022) in descending order. We selected the top 10 in this list, as shown in \autoref{tab:system_selection}. 

\begin{table}[ht]
    \centering
    \caption{Overview of the selected GitHub repositories.}
    \begin{tabular}{@{}lllrr@{}}
    \toprule
    \textbf{Repository} & \textbf{Branch}      & \textbf{Commit} & \textbf{Forks + Stars} & \textbf{\phantom{.} Files (.h, .cpp)} \\ \midrule
    UnrealEngine & v4 & 90f6542cf7       & 64100 & 66390       \\
    godot      & 3.4 & f9ac000d5d      & 59200  & 5603      \\
    cocos2d-x    & v4 & 90f6542cf7      & 23300  & 1601      \\
    o3de       & development  & 21ab0506da      & 6400   & 7278      \\
    Urho3d     & master     & feb0d90190      & 4956   & 4312      \\
    gamePlay3d & master     & 4de92c4c6f      & 4900   & 688      \\
    panda3D    & master     & 2208cc8bff      & 4100  & 5344       \\
    olcPixelGameEngine   & master     & 02dac30d50      & 3963  & 81       \\
    Piccolo    & main       & b4166dbcba      & 3892  & 1572       \\
    FlaxEngine  & master     & 7b041bbaa5      & 3613  & 2134       \\ \bottomrule
    \end{tabular}
    \label{tab:system_selection}
\end{table}

\subsection{Subsystem Selection}
\label{sec:subsystem_selection}
We use the 15 subsystems described in our reference architecture, the ``Runtime Engine Architecture'' (\autoref{fig:gregory_arch}). Given that commercial game engines provide an integrated development environment, we added the ``World Editor'' subsystem in our analysis, totalling 16 subsystems.

For brevity, we identify subsystems in the reference architecture with 3-letter identifiers: \textit{Audio} (AUD), \textit{Core} (COR), \textit{Profiling and Debugging} (DEB), \textit{Front End} (FES), \textit{Gameplay Foundations} (GMP), \textit{Human Interface Devices} (HID), \textit{Low-Level Renderer} (LLR), \textit{Online Multiplayer} (OMP), \textit{Collision and Physics} (PHY), \textit{Platform Independence Layer} (PLA), \textit{Resources} (RES), \textit{Third-party SDKs} (SDK), \textit{Scene graph/culling optimizations} (SGC), \textit{Skeletal Animation} (SKA), \textit{Visual Effects} (VFX), \textit{World Editor} (EDI).

\subsection{Subsystem Detection}
\label{sec:subsystem_detection}

In this step, we clustered all folders in each repository into the selected subsystems. When deciding which folders belong to a subsystem, we considered four pieces of information from each folder: its name, contents, documentation, and source code. We show an example of this decision process in \autoref{tab:subsystem_detection}.

\begin{table}[ht]
\centering
\caption{Subsystem detection example for Cocos2d-x.}
\label{tab:subsystem_detection}
\begin{tabular}{@{}p{3.6cm}p{8cm}@{}}
\toprule
\multicolumn{2}{l}{\textbf{Can we determine the subsystem of /cocos/editor{\textendash}support/spine by:}} \\ \hline
1) Folder name? & No, the name \textit{spine} does not match or relate to the reference architecture. \\
2) Parent folder name? & No, the folder \textit{editor-support} might be related to EDI, but we need more data to confirm. \\
3) Documentation?      & \textbf{Yes}, according to docs: ``Skeletal animation assets in Creator are exported from Spine''. \tablefootnote{\url{https://docs.cocos.com/creator/manual/en/asset/spine.html}} \\
4) Source code? & No code analysis needed, subsystem detected on step 3. \\ \hline
\textbf{Conclusion}    & \textit{Skeletal Animation} (SKA)   \\ \hline
\end{tabular}
\end{table}

\subsection{Include Graph Generation}
\label{sec:include_graph_gen}

In parallel to detecting subsystems, we generated an \textit{include} graph of each game engine using a two-pass algorithm. In the first pass, our analyser reads every source code file composing the game engine, collects all includes and outputs an \textit{include} graph in the DOT graph description language. In the output DOT file, each row is an  \textit{include} relationship described as follows: \textit{/home/engine/source.cpp -$>$ /home/engine/target.h}. The analyser attempts to resolve each relative \textit{include} path into an absolute path. If the resolution fails, the analyser writes the path to another file called \textit{engine-includes-unr.csv}. 

In the original implementation of the approach \cite{ullmann_gas_2023}, we read and resolved each of the unresolved \textit{include} paths manually. However, repeating this operation for thousands of paths is time-consuming and error-prone. Therefore, we automated this step by adding a second pass to our analyser. In this pass, it loads \textit{engine-includes-unr.csv}, iterates over each of its paths, and splits them by their folder delimiters. Then it searches for each part of the path, starting with the file name and moving towards the repository root folder. It repeats this search until it finds a match. Finally, the resolved absolute path is appended to the DOT file.

Some \textit{include} paths inevitably remain unresolved because they refer to system or OS-specific libraries (e.g., \textit{stdio.h}, \textit{windows.h}) which do not belong to the game engine. In Cocos2d-x, all third-party dependencies are located in a separate repository. However, these paths do not contain code written by game engine developers, so their absence is not detrimental to the consistency of our architectural models.

\subsection{Moose Model Generation}
\label{sec:moose_model_gen}

In this step, we merge the data collected in step 3 (the CSV file containing the detected subsystems) and step 4 (the DOT file containing the \textit{include} graph) for each game engine. We build on Moose 10 \footnote{\url{https://github.com/moosetechnology}}, a platform for software analysis implemented in Pharo \footnote{\url{https://pharo.org/}}, a Smalltalk development environment. By importing the files into Moose, we create a Moose model, which is our architectural model.

\subsection{Architectural Model Visualisation}
\label{sec:architectural_model_gen}

Finally, we use Moose's ``Architectural map'' visualisation, which displays all subsystems and their relationships, to visualise each of the 10 selected game engines. By compiling the information from  all generated ``Architectural maps'', we created a heatmap which we will show and explain in \autoref{sec:rq2}. We also used Gephi \footnote{\url{https://gephi.org/}}, a graph analysis software, to compute metrics such as in-degree and betweenness centrality, which are the main point of discussion in \autoref{sec:rq1}. We chose to use Gephi instead of Moose in this case because Moose does not come with graph analysis tools out of the box.

\section{Results}
\label{sec:results}

\begin{figure}[ht]
  \centering
  \begin{subfigure}{0.49\linewidth}
    \includegraphics[width=\linewidth]{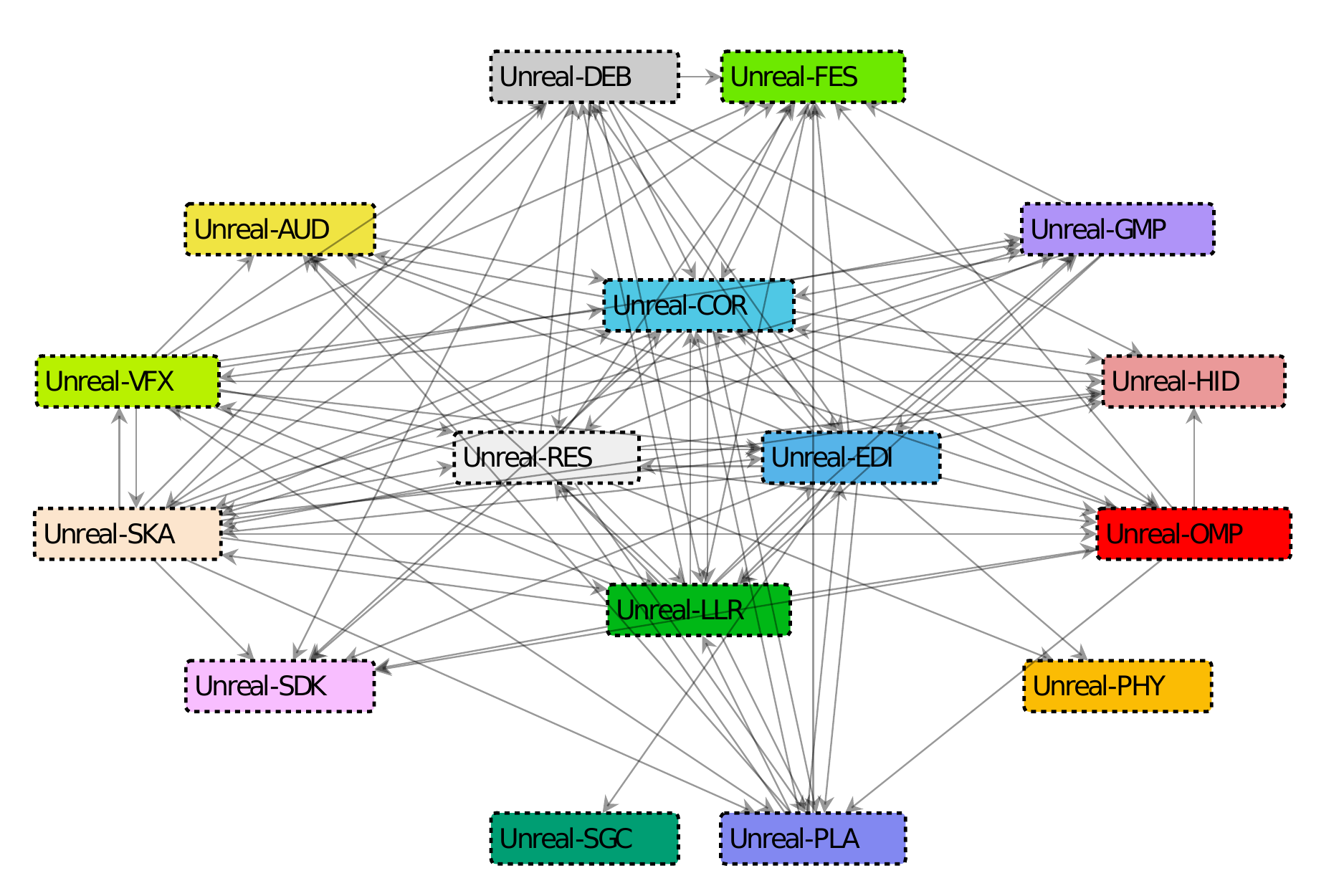}
    \caption{Unreal Engine}
    \label{fig:results_unreal}
  \end{subfigure}
  \hfill
  \begin{subfigure}{0.47\linewidth}
    \includegraphics[width=\linewidth]{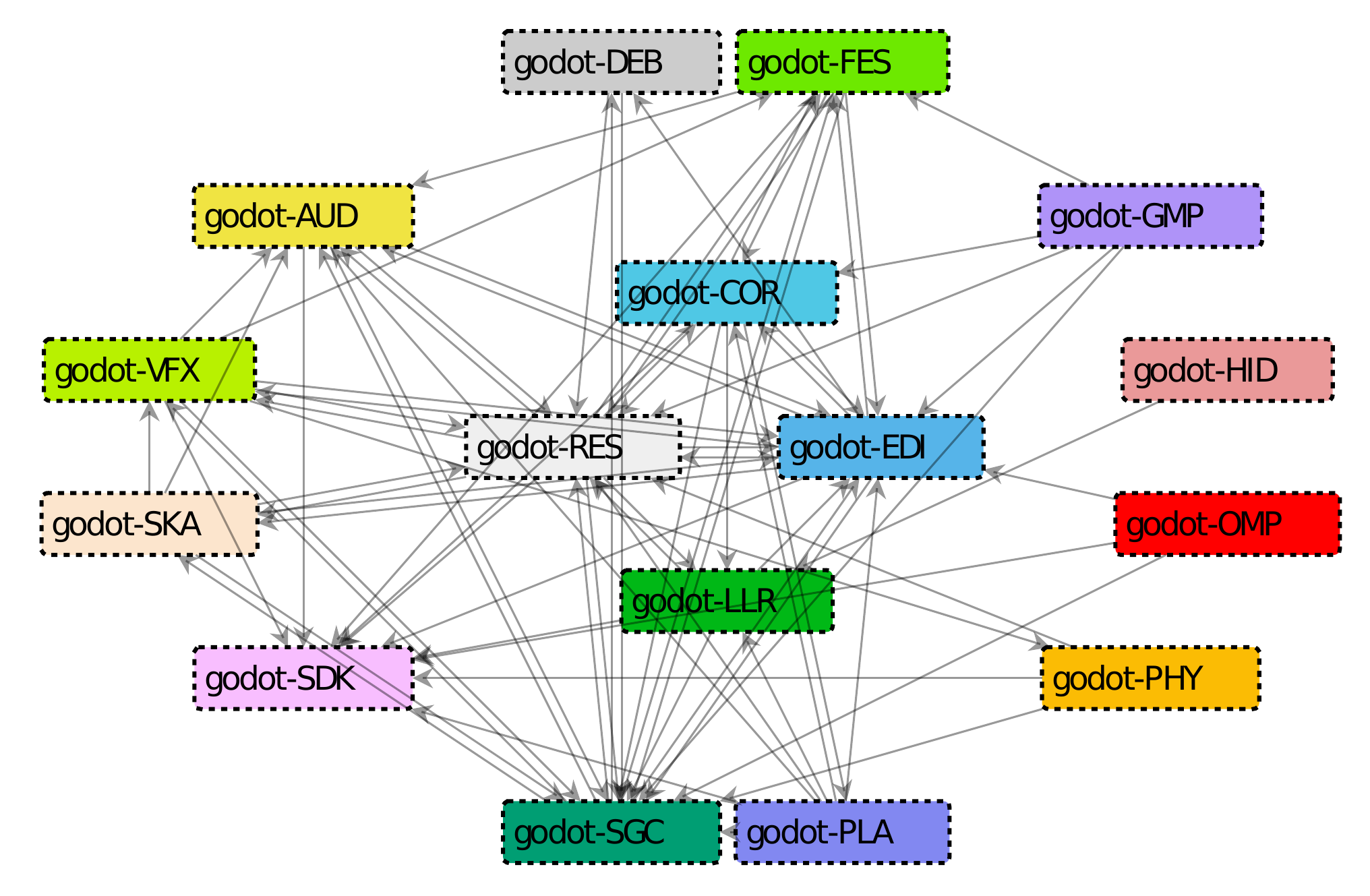}
    \caption{Godot}
    \label{fig:results_o3de}
  \end{subfigure}
  \caption{Game engine architectural models generated with Moose 10.}
  \label{fig:arch_maps}
\end{figure}

\autoref{fig:arch_maps} shows the architectural models of Godot and Unreal Engine. The high number of relationships and subsystems shows that both game engines are highly coupled and follow the reference architecture. In Godot, the most coupled subsystem is \textit{Scene graph/culling optimizations} (SGC) because it centralizes files with diverse functionalities in its \textit{/scene/3d} folder, such as \textit{camera.h} for \textit{Low-Level Renderer} (LLR) and \textit{physics\_body.h} for \textit{Collision and Physics} (PHY). In contrast, in Unreal's \textit{/Engine/Source/Runtime} folder, we observe LLR and PHY are divided in several distinct subfolders (e.g. \textit{PhysicsCore}, \textit{Renderer}).

While the ``Architectural map'' provides us with an overview of coupling in a given game engine, its density makes interpretation hard, especially because we want to understand how two or more subsystems are coupled. In the following subsections, we answer our RQs and explain how subsystem coupling can be identified and understood with the use of graph analysis and a coupling heatmap.

\subsection{RQ1 - Which Subsystems More Often Couple with One Another?} 
\label{sec:rq1}

\begin{figure}[ht]
    \centering
    \includegraphics[width=\linewidth]{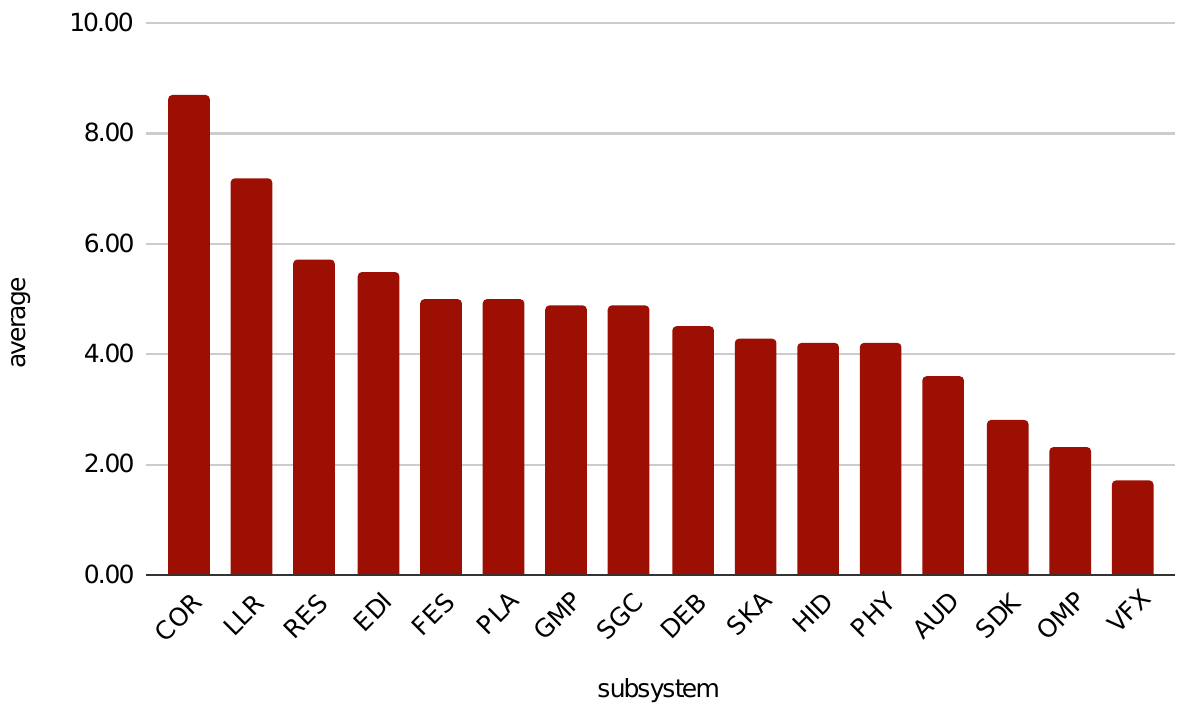}
    \caption{Average subsystem in-degree.}
    \label{fig:results_indegree}
\end{figure}

We computed the in-degree for each subsystem of each game engine. Next, we computed the averages and sorted them in descending order. As we can observe in \autoref{fig:results_indegree}, the top-five subsystems in average in-degree are: \textit{Core} (COR), \textit{Low-Level Renderer} (LLR), \textit{Resources} (RES), \textit{World Editor} (EDI) and, tied in 5th place, \textit{Front End} (FES) and \textit{Platform Independence Layer} (PLA). 

The subsystems in the top-five act as a foundation for game engines because most of the other subsystems depend on them to implement their functionalities. Two subsystems in this list are graphics-related: \textit{Low-Level Renderer} (LLR) and \textit{Front End} (FES). We expected it because video games depend on visuals.

\begin{figure}
    \includegraphics[width=\linewidth]{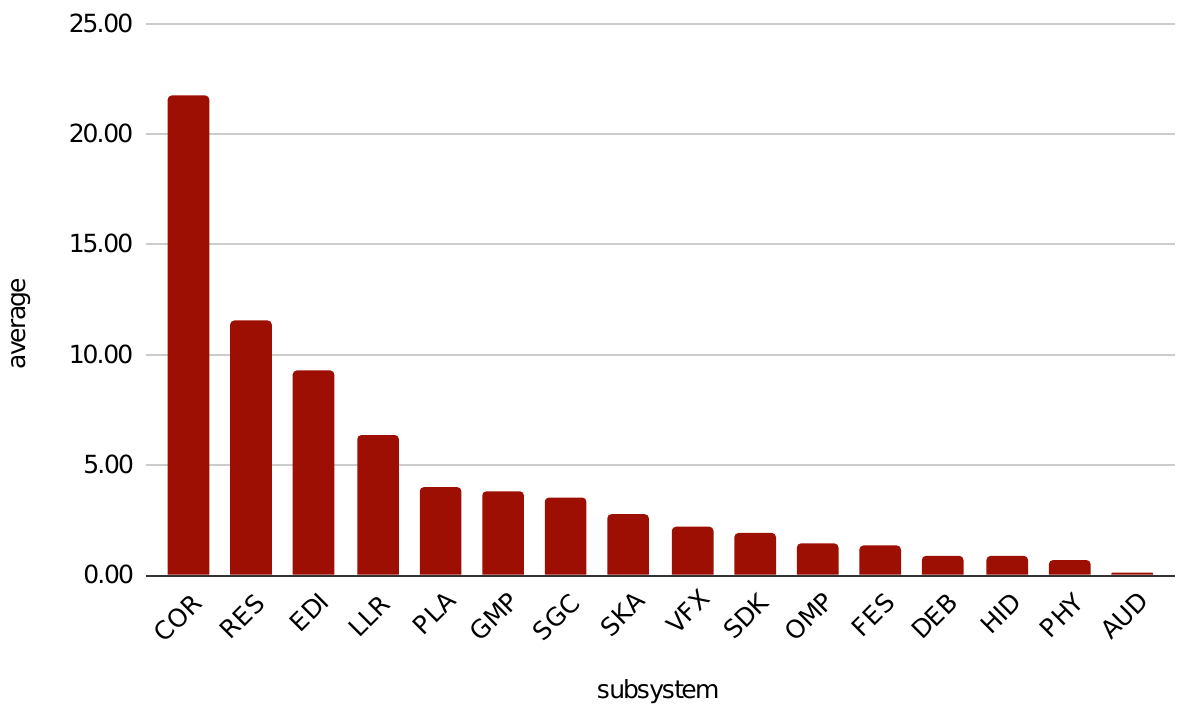}
    \caption{Average subsystem betweenness centrality.}
    \label{fig:results_centrality}
\end{figure}

Same as the in-degree, we computed the average betweenness centrality, shown in \autoref{fig:results_centrality}. This metric represents the extent to which a node lies in the path of others \cite{badar_examining_2013}. It helps us understand whether a highly coupled subsystem is an isolated occurrence, or if it consistently plays a central role within its respective system. \autoref{fig:results_centrality} shows that the top-five systems with the highest average betweenness centrality are: \textit{Core} (COR), \textit{Resources} (RES), \textit{World Editor} (EDI), \textit{Low-Level Renderer} (LLR) and \textit{Platform Independence Layer} (PLA). For this reason, we decided to draw the top four subsystems in the centre of the architectural maps shown in \autoref{fig:arch_maps}. 

We observe that all subsystems with high in-degree also have high centrality, being \textit{Front End} (FES) the only exception. While subsystems frequently depend on FES, we observe FES often depends only on \textit{Core} (COR) and \textit{Low-Level Renderer} (LLR) and therefore does not play the role of intermediate or ``gatekeeper'' between groups of subsystems. We further discuss subsystem ``gatekeeping'' observed in the COR subsystem in \autoref{sec:rq2}.

\subsection{RQ2 - Do Game Engines Share Subsystem Coupling Patterns?}
\label{sec:rq2}

\begin{figure}[ht]
    \centering
    \includegraphics[width=\textwidth]{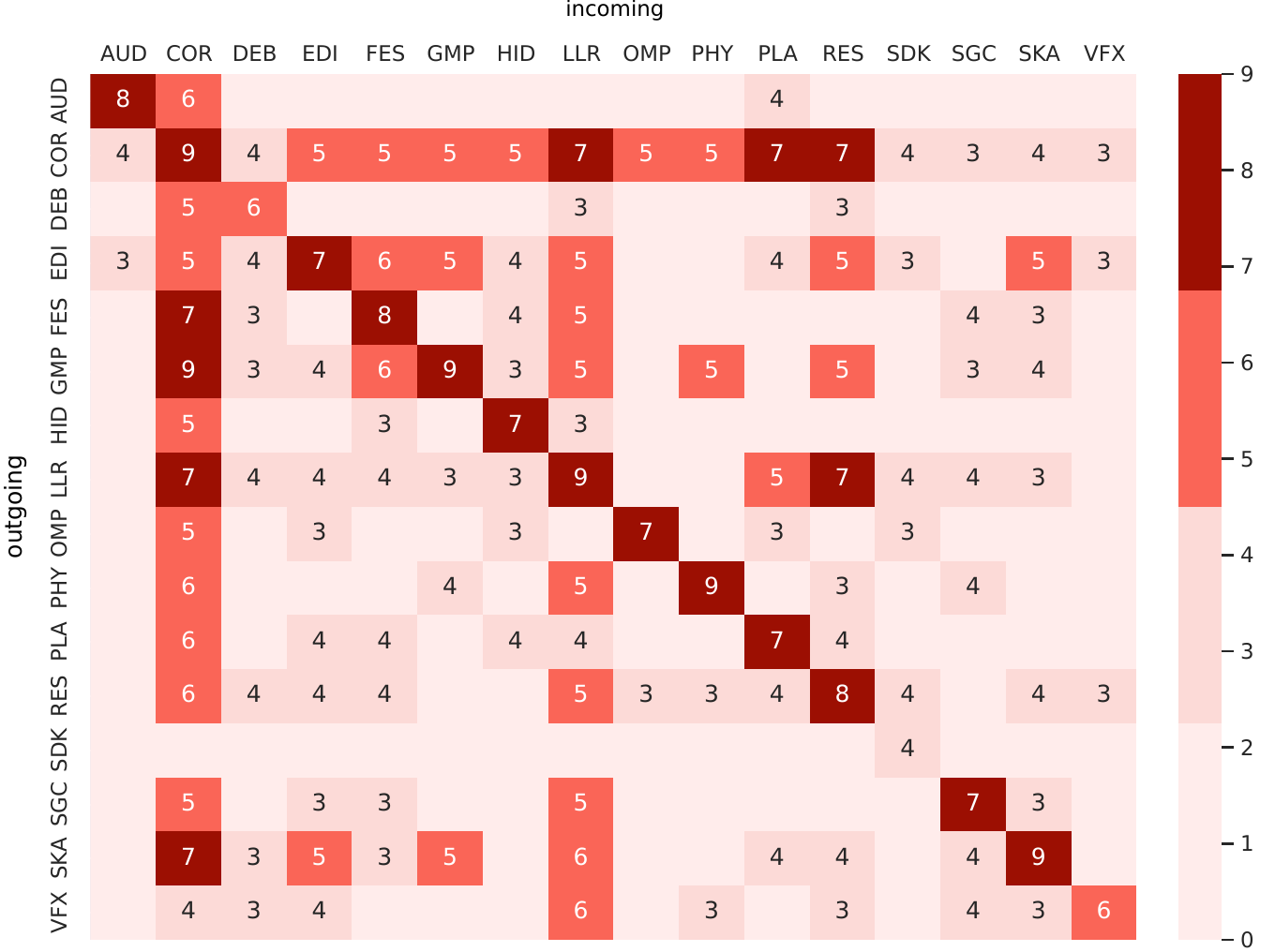}
    \caption{Subsystem coupling heatmap showing aggregated coupling counts.}
    \label{fig:results_heatmap}
\end{figure}

To answer this question, we created a heatmap which shows aggregated coupling counts from all architectural models (\autoref{fig:results_heatmap}). For example, if we take the first line from the top, we can observe the \textit{Audio} (AUD) subsystem includes files from itself in eight game engines, and it includes files from COR in six game engines. We show the most frequent coupling pairs from the heatmap in \autoref{tab:results_frequent_coupling}.

While we applied our approach to 10 game engines, no square shows the value 10 in the heatmap's central diagonal. This happens because olcPixelGameEngine is fully decoupled. As an educational game engine, each of its subsystems was written in a single .h file, which is meant to be included by developers to their own .cpp file. Also, not all subsystems were detected in all game engines, and therefore not all self-include nine times.

\begin{table}[ht]
\centering
\caption{The most frequent subsystem coupling pairs.}
\label{tab:results_frequent_coupling}
\begin{tabular}{lclrc|clclrc|clclr}
\hline
\textbf{Pair} & \textbf{} & \textbf{} & \textbf{Count} & & & \textbf{Pair} & & & \textbf{Count} & & & \multicolumn{1}{l}{\textbf{Pair}} & \multicolumn{1}{l}{} & & \multicolumn{1}{r}{\textbf{Count}} \\ \hline
GMP & -\textgreater{} & COR & 9 & & & FES & -\textgreater{} & COR & 7 & & & EDI & -\textgreater{} & FES & 6 \\
COR & -\textgreater{} & LLR & 7 & & & LLR & -\textgreater{} & COR & 7 & & & GMP & -\textgreater{} & FES & 6 \\
COR & -\textgreater{} & PLA & 7 & & & SKA & -\textgreater{} & COR & 7 & & & PLA & -\textgreater{} & COR & 6 \\
COR & -\textgreater{} & RES & 7 & & & AUD & -\textgreater{} & COR & 6 & & & RES & -\textgreater{} & COR & 6 \\ \bottomrule
\end{tabular}
\end{table}

While the analysis of in-degree and betweenness centrality highlights which subsystems are fundamental, the heatmap shows how they work together. \textit{Core} (COR) is the subsystem that most frequently includes others, and also the most frequently included. It is often reciprocally related to \textit{Resources} (RES) and \textit{Platform Independence Layer} (PLA), reflecting a ``gatekeeper'' role described in the reference architecture: when loading or saving game assets, RES uses PLA to interface with the OS and hardware such as the hard disk. We will further explore these frequent relationships in \autoref{sec:discussion}.

\section{Discussion}
\label{sec:discussion}

\begin{figure}[ht]
    \centering
    \includegraphics[width=0.8\textwidth]{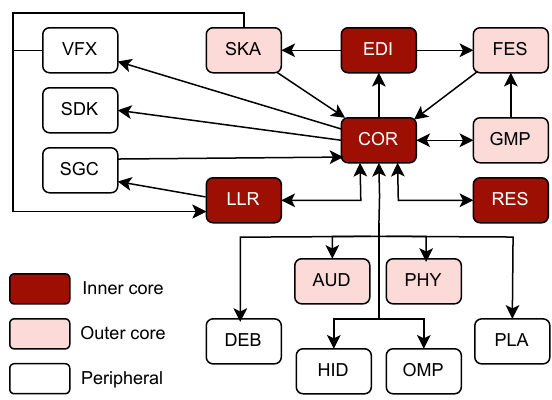}
    \caption{Our emergent open-source game engine architecture.}
    \label{fig:emergent_arch}
\end{figure}

By compiling the game engine coupling pattern information from \autoref{tab:results_frequent_coupling} we observe a new architecture emerge. In \autoref{fig:emergent_arch}, we placed in the centre of the model the subsystems with the highest betweenness centrality, forming an inner core (dark red). Next, we placed other subsystems which appear in \autoref{tab:results_frequent_coupling} in the outer core (light red). Finally, we placed the subsystems which do not appear in \autoref{tab:results_frequent_coupling} in the outer core's periphery (white). All relationships shown in the diagram are among the most frequent, as shown in \autoref{tab:results_frequent_coupling}  and \autoref{fig:results_heatmap}. When there was a tie (e.g. two pairs had the same frequency), we chose the coupling pair with the highest sum of betweenness centrality. 

In this emergent architecture, we observe the \textit{Low-Level Renderer} (LLR) often inter-depends on \textit{Core} (COR), which it uses to access functionality in the \textit{Platform Compatibility Layer} (PLA) and the \textit{Resources} (RES) subsystem. In \autoref{fig:gregory_arch}, we can observe these subsystems are also placed close to each other in the reference architecture.

While not part of the inner core, the \textit{Front End} (FES) subsystem plays an important role. It is often included by the \textit{World Editor} (EDI) and \textit{Gameplay Foundations} (GMP), which are both visual interfaces between the user and the game engine. Because it manages UI elements which emit events and trigger actions throughout the system, \textit{Front End} (FES) often depends on the event/messaging system in \textit{Core} (COR).

More practically, the information provided by the architectural models and coupling patterns can be used by practitioners as follows:
\begin{itemize}
    \item \textbf{Learning}: architectural model visualisations provide a friendly way for novice game engine developers to understand this kind of system and start developing their own subsystems or plugins.
    \item \textbf{Refactoring}: game engine developers can refactor their code more safely by visualising how changes to a subsystem could impact the whole game engine.
    \item \textbf{Anomaly Detection}: a subsystem coupling heatmap can help game engine developers to find unusual or unexpected coupling patterns, and then improve the source code as necessary.
    \item \textbf{Reference Extraction}: game engine architects seeking to design a new engine can extract architectural models from similar systems and use them as references. This is useful both for large companies and small indie developers who develop tailor-made solutions, e.g. for performance.
\end{itemize}

\section{Threats to Validity}
\label{sec:threats}

First, the selected game engines may not be representative of all open-source game engines and the entire video game industry. Similarly, \citeauthor{gregory_game_2018} \cite{gregory_game_2018} is our reference architecture and we are aware other architectures exist, as explained in \autoref{sec:related_work}, even though not as detailed. Moreover, we acknowledge some modern game engine features, such as AR and VR support, were not present in the subsystem selection because they are not described in the reference architecture.

Subsystem detection was performed manually by the first author only, which may bias the detection process. To mitigate this issue, we intend to assign multiple people to work in this step and later combine the results by consensus. We are also aware that our approach is dependent on the behaviour and metrics provided by Moose and Gephi, and changing them could also change the results and therefore our perception of these game engine architectures.

\section{Conclusion}
\label{sec:conclusion}

In this paper, we show that by generating and studying game engine architectural models, game engine developers can identify which subsystems are the centrepieces of their system and therefore give them proper maintenance. Additionally, by understanding frequent coupling patterns, game engine architects can take better decisions when extending existing game engines or creating custom-made solutions for a specific kind of video game or interactive experience.

In future work, we will apply our approach to a wider variety of game engines and subsystems. We will also conduct experiments with developers to determine to which extent the visualisations produced by our approach can help improve system understanding and maintainability. Finally, we intend to explore automated approaches to architecture recovery and other software quality metrics, such as cohesion and complexity.

\section*{Acknowledgements}
The authors were partially supported by the NSERC Discovery Grant and Canada Research Chairs programs.

\bibliographystyle{plainnat}
\bibliography{main.bib}
\end{document}